\documentclass[pra,aps,showpacs,twocolumn,superscriptaddress,amsmath,amssymb]{revtex4}

\usepackage{multirow}
\usepackage{graphicx,dcolumn,bm}
\usepackage{mathrsfs}
\usepackage[usenames,dvipsnames]{color}
\usepackage{subfigure}
\newcommand {\startbild}{\begin{figure}}
\newcommand {\stopbild}{\end{figure}}

\newcommand {\staf}{\begin{equation}}
\newcommand {\stof}{\end{equation}}
\newcommand {\staffeld}{\begin{eqnarray}}
\newcommand {\stoffeld}{\end{eqnarray}}

\newcommand{\ket}[1]{|#1\rangle}
\newcommand{\bra}[1]{\langle #1|}

\begin{document}

\title{Quantum interference initiated super- and subradiant emission from entangled atoms}

\author{R. Wiegner}
\email{ralph.wiegner@physik.uni-erlangen.de}
\affiliation{Institut f\"ur Optik, Information und Photonik, Universit\"at Erlangen-N\"urnberg, Erlangen, Germany}
\homepage{http://www.ioip.mpg.de/jvz/}

\author{J. von Zanthier}
\affiliation{Institut f\"ur Optik, Information und Photonik, Universit\"at Erlangen-N\"urnberg, Erlangen, Germany}
\affiliation{Erlangen Graduate School in Advanced Optical Technologies (SAOT), Friedrich-Alexander-Universit\"at Erlangen-N\"urnberg}

\author{G. S. Agarwal}
\affiliation{Department of Physics, Oklahoma State University, Stillwater, OK, USA}

\date{\today}

\begin{abstract}
We calculate the radiative characteristics of emission from a system of entangled atoms which can have a relative distance larger than the emission wavelength. We develop a quantum multipath interference approach which explains both super- and subradiance though the entangled states have zero dipole moment. We derive a formula for the radiated intensity in terms of different interfering pathways. We further show how the interferences lead to directional emission from atoms prepared in symmetric W-states. As a byproduct of our work we show how Dicke's classic result can be understood in terms of interfering pathways. In contrast to the previous works on ensembles of atoms, we focus on finite numbers of atoms prepared in well characterized states.
\end{abstract}

\maketitle

\section{Introduction}

The phenomenal progress in the preparation of entangled states of atoms, particularly in chains of trapped ions \cite{BlattWineland08,Blatt04b,Blatt05,Wineland05,Blatt11}, has enabled one to demonstrate many basic tasks required for quantum computation \cite{Cirac95,Cirac03}. It has been realized and demonstrated that entangled states also provide one with precision methods for doing quantum metrology \cite{Wineland05,Dowling02,Leibfried04,Lloyd04,Dowling08,Gerry10}. However it has been much less conceived that entangled states give us a new paradigm for doing optical physics \cite{Agarwal11} which traditionally is done using independent atoms though with exceptions \cite{Dicke54,Scully06,Scully07,Scully09,Scully09b,Scully10,Cirac11}. Since one has succeeded in preparing well characterized entangled states albeit for a small number of qubits, it is pertinent to ask how optical effects can depend on both the nature of the entangled states as well as on the number of atoms. In particular it is pertinent to ask how the radiative properties of atoms in well characterized entangled states differ from those of atoms prepared in separable states. The simplest system to study is a system of two two-level atoms prepared in an entangled state and this has been extensively studied for its dynamical evolution \cite{Brewer96,Ficek02,Eberly04,Agarwal06,Agarwal08}. 

In this paper we examine a system of N atoms prepared in well characterized entangled states like W-states where the interatomic distance is larger than the emission wavelength. We show how the nature of the initial W-state  dictates its radiative characteristics leading to superradiant emission of photons. It must be added that superradiance has been studied extensively since its prediction by Dicke \cite{Dicke54}. Much of the literature deals with ensembles of atoms with inherent complexities associated with ensembles. In contrast we deal with a finite number of atoms prepared in well characterized entangled states. This enables us to give a very clear physical picture based on the interference of quantum paths. Note that if the system has a finite dipole moment then we can easily interpret superradiance as due to a large macroscopic dipole moment. However for atoms in W-states there is no macroscopic dipole moment and the standard argument cannot be used for the occurence of superradiance. We also note that most of the work on entangled states is driven by quantum computation protocols. Thus the idea of doing optical physics with entangled atoms should give a new impetus to the generation of entangled states for large numbers of atoms or of other quantum systems displaying a similar behavior \cite{Wrachtrup08,Lukin10}.

The paper is organized as follows: In section \ref{enhancedW} we investigate the enhancement in the emission of radiation scattered by atoms which are 
initially prepared in generalized symmetric W-states. We next examine the physics behind such an enhancement. We find that the enhancement can be explained by an interference of multiple photon quantum paths. We develop a framework which enables to calculate the number of quantum paths and the contribution of each quantum path. In section \ref{direct} we investigate the angular dependence of photons emitted by the entangled system and give explicit results for any number of atoms. Finally, in section \ref{antiW}, we extend our multipath quantum interference approach for radiation from non-symmetric generalized W-states.

\section{Enhanced emission from arbitrary symmetric W-states}
\label{enhancedW}

\begin{figure}[h!]
\centering
\includegraphics[width=0.4 \textwidth]{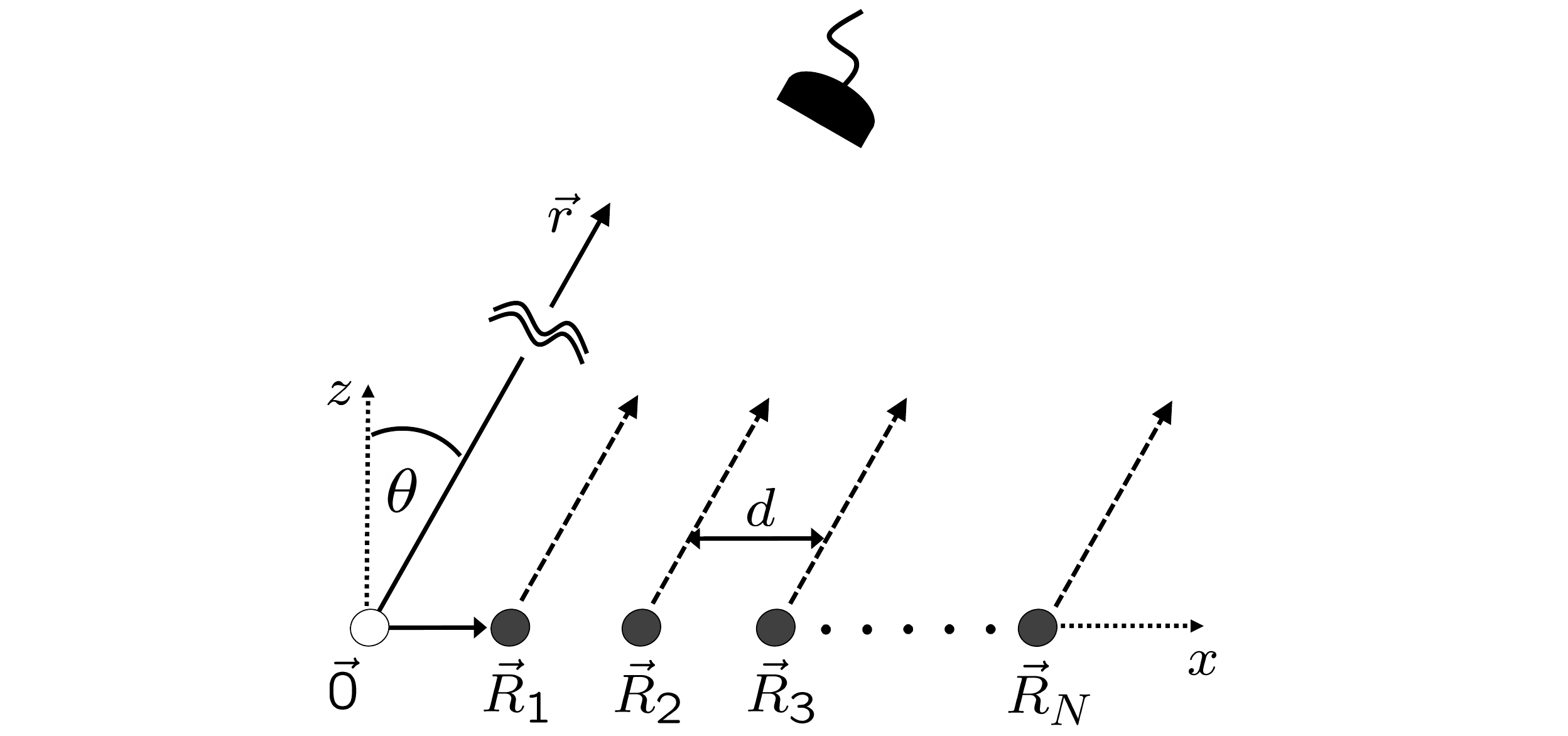}
\caption{\label{setup} Scheme of the investigated system: $N$ equidistant two-level atoms are localized along a chain at positions ${\vec R}_1$ to ${\vec R}_N$. A detector at position ${\vec r}$ registers a photon, scattered by the atoms, in the far field.}
\end{figure}

In the following we assume a linear chain of $N$ identical two-level atoms with upper level $\ket{e}$ and ground state $\ket{g}$ localized at positions ${\vec R}_1\,, ...\,,{\vec R}_N$, where for simplicity we consider an equal spacing $d$ between adjacent atoms (cf.~Fig.~\ref{setup}). We assume $k\,d > 1$ so that the dipole-dipole interaction can be neglected, where $k=\frac{2\pi}{\lambda}$ denotes the wave number of the transitions $\ket{e} \rightarrow \ket{g}$. The initial state of the chain of atoms is taken to be a symmetric W-state. We can construct W-states by assuming, say, one atom is excited and the rest of the atoms is in the ground state. This state would then be of the form

\staffeld
\ket{W} &=& \frac{1}{\sqrt N} (\ket{e\,g\,...\,g} + \ket{g\,e\,g\,...\,g} + \ldots + \ket{g\,...\,g\,e}) \nonumber\\
 &=& \ket{W_{1,N-1}}\,.\hspace{2cm}
\stoffeld

\noindent The latter notation implies that one atom is excited and $(N-1)$ atoms are in the ground state. We will see that the radiative properties of the W-state are quite different from the properties of a separable state like $\ket{e\,g\ldots\,g}$. The entanglement in the W-state endows it with characteristic radiative properties. We can also consider more general W-states $\ket{W_{n_e,N-n_e}}$, where $n_e$ atoms are excited and $N-n_e = n_g$ atoms are in the ground state

\staf
\label{W-state}
\ket{W_{n_e,N-n_e}} \equiv  \binom{N}{n_e}^{-\frac{1}{2}} \sum_k {\bf P}_k \ket{S_{n_e,N-n_e}}\,.
\stof

\noindent Here, $\{{\bf P}_k\}$ denotes the complete set of all possible distinct permutations of the qubits. For example, for $n_e = 2$ and $N - n_e = 2$ the state of Eq.~(\ref{W-state}) would take the explicit form

\staffeld
\label{WEXAMPLE2}
\ket{W_{2,2}} = \frac{1}{\sqrt{6}} \left(\ket{e\;e\;g\;g} + \ket{e\;g\;e\;g} + \ket{e\;g\;g\;e}\right.\nonumber\\
\left.+ \ket{g\;e\;e\;g}+ \ket{g\;e\;g\;e}+ \ket{g\;g\;e\;e} \right)\,.
\stoffeld

\noindent Further we introduce the separable states whose radiative properties we will compare to the radiative properties of the W-states in the following, defined by

\staf
\label{S-state}
\ket{S_{n_e,N-n_e}} \equiv \prod_{\alpha = 1}^{n_e} \ket{e_\alpha} \prod_{\beta = n_e+1}^{N} \ket{g_\beta}\,.
\stof

We now consider a detector placed at position ${\vec r}$ in the far-field region of the atoms to measure the intensity

\staf
\label{Eq1}
\mathrm{I} = \langle \hat{E}^{(-)}\, \hat{E}^{(+)} \rangle
\stof

\noindent emitted by the atomic chain. The positive frequency part of the electric field operator is given by \cite{Agarwal74} 

\staf
\label{Eq2a}
\hat{E}^{(+)} \sim \frac{e^{i\,k\,r}}{r} \sum_j \vec{n} \times\,(\vec{n} \times \vec{p}_{ge}) \, e^{-i\,\varphi_j} \;\hat{s}^{-}_j \,,
\stof

\noindent with the unit vector $\vec{n} = \frac{\vec{r}}{r}$ and $\vec{p}_{ge}$ the dipole moment of the transition $\ket{e}\rightarrow\ket{g}$. Furthermore, we denote with $\hat{s}^{-}_j = |g\rangle_j\langle e|$ the dipole operator and with $\varphi_j$ the relative optical phase accumulated by a photon emitted at $\vec{R}_j$ and detected at ${\vec r}$, where (cf.~Fig.~\ref{setup})

\staf
\label{Eq3}
\varphi_j({\vec r}) \equiv \varphi_j = k \,\vec{n}\cdot\vec{R}_j = j\,k\,d\,\sin\theta\,.
\stof

\noindent The negative frequency part of the electric field operator is obtained by Hermitian conjugation, i.e., \mbox{$\hat{E}^{(-)} = \hat{E}^{{(+)}^\dagger}$}. In the following we will consider for reasons of clarity $\vec{p}_{ge}$ to be along the $y$ direction and $\vec{n}$ in the $x-z$ plane, so that $\vec{p}_{ge}\cdot \vec{n} = 0$ . With these assumptions and omitting all constant factors Eq.~(\ref{Eq2a}) simplifies to
\staf
\label{Eq2}
\hat{E}^{(+)} \sim \sum_j e^{-i\,\varphi_j} \;\hat{s}^{-}_j\,.
\stof

\noindent The field is now dimensionless and hence all intensities would be dimensionless. The actual values can be obtained by multiplying the emission produced by a single excited atom. Eq.~(\ref{Eq2}) leads to the following expression for the radiated intensity at ${\vec r}$ (cf., Eq.~(\ref{Eq1}))

\staffeld
\label{IGEN}
\mathrm{I}({\vec r}) &=& \sum_{i,j} \langle \hat{s}^{+}_i \hat{s}^{-}_j \rangle \,e^{i\,(\varphi_i-\varphi_j)}\nonumber\\
&=& \sum_{i} \langle \hat{s}^{+}_i \hat{s}^{-}_i \rangle + \left(\sum_{i\neq j} \langle \hat{s}^{+}_i \rangle\langle\hat{s}^{-}_j \rangle \right.\nonumber \\
&&\hspace{0.5cm}\left. + \sum_{i\neq j} (\langle \hat{s}^{+}_i \hat{s}^{-}_j \rangle - \langle \hat{s}^{+}_i \rangle\langle\hat{s}^{-}_j \rangle )\right)e^{i\,(\varphi_i-\varphi_j)}\,.
\stoffeld

\noindent Thus the characteristics of the intensity would depend on the incoherent terms $\langle \hat{s}^{+}_i \hat{s}^{-}_i \rangle$, the nonvanishing of the dipole moment $\langle \hat{s}^{+}_i\rangle$ and quantum correlations like $\langle \hat{s}^{+}_i \hat{s}^{-}_j \rangle - \langle \hat{s}^{+}_i \rangle\langle\hat{s}^{-}_j \rangle$. In case of the $N$-qubit separable state $\ket{S_{n_e,N-n_e}}$ the intensity calculates to

\staf
\label{ISMAX}
\mathrm{I}_{\ket{S_{n_e,N-n_e}}} = \sum_{i,j = 1}^{n_e} \langle \hat{s}^{+}_i \hat{s}^{-}_j \rangle \,e^{i\,(\varphi_i-\varphi_j)}= \sum_{i= 1}^{n_e} \langle \hat{s}^{+}_i \hat{s}^{-}_i \rangle = n_{e}\,.
\stof

\noindent Here, we have explicitly used that for separable states we find

\staf
\langle \hat{s}^{+}_i \hat{s}^{-}_j \rangle = \langle \hat{s}^{+}_i \rangle \langle \hat{s}^{-}_j \rangle = 0,\,\mathrm{for}\; i \neq j\,,
\stof
 
\noindent since $\langle \hat{s}^{+}_j \rangle = 0$, i.e., the dipole moment $\langle \hat{s}^{+}_j \rangle$ as well as the correlations $\langle \hat{s}^{+}_i \hat{s}^{-}_j \rangle$ for $i \neq j$ are zero. According to Eq.~(\ref{ISMAX}) the intensity distribution of separable states is a constant corresponding simply to the number of initially excited atoms, i.e., $\mathrm{I}_{\ket{S_{n_e,N-n_e}}} = \mathrm{I}_{\ket{S_{n_e,0}}}$ for any $N$. This can be explained as every atom radiates incoherently. Note that in case of the celebrated realization of Young's double slit experiment using independent atoms coherently excited by a cw laser \cite{Wineland93} the quantum correlations in Eq.~(\ref{IGEN}) are zero. However, the dipole moment $\langle \hat{s}^{+}_j \rangle$ is nonzero, whence Eq.~(\ref{IGEN}) reduces to

\staffeld
\label{IGEN5}
\mathrm{I}({\vec r}) = \sum_{i} \langle \hat{s}^{+}_i \hat{s}^{-}_i \rangle + \sum_{i\neq j} \langle \hat{s}^{+}_i \rangle\langle\hat{s}^{-}_j \rangle \,e^{i\,(\varphi_i-\varphi_j)}\,,
\stoffeld

\noindent which leads to interferences in the mean radiated intensity.

Before we derive the intensity distribution of the generalized W-states $\ket{W_{n_e,N-n_e}}$, and thus the enhancement in the emission of radiation scattered by atoms which are initially prepared in these states, we will illustrate our key ideas with a simple example. Consider a system of three atoms prepared in the W-state 

\staf
\label{WEXAMPLE1}
\ket{W_{1,2}} = \frac{1}{\sqrt{3}} \left(\ket{e\,g\,g} + \ket{g\,e\,g} + \ket{g\,g\,e} \right)\,.
\stof

\noindent The intensity of this state calculates to

\staffeld
\label{IW12}
\mathrm{I}_{\ket{W_{1,2}}} &=& \sum_{i,j = 1}^{3} \langle \hat{s}^{+}_i \hat{s}^{-}_j \rangle \,e^{i\,(\varphi_i-\varphi_j)} \nonumber\\
&=&\sum_{i= 1}^{3} \langle \hat{s}^{+}_i \hat{s}^{-}_i \rangle + \sum_{\substack{i,j=1\\i \neq j}}^3 \langle \hat{s}^{+}_i \hat{s}^{-}_j \rangle \,e^{i\,(\varphi_i-\varphi_j)}\,,
\stoffeld

\noindent with the dipole moment $\langle \hat{s}^{+}_j \rangle$ being again zero. The first sum $\langle \hat{s}^{+}_i \hat{s}^{-}_i \rangle$ in Eq.~(\ref{IW12}) corresponds to the intensity of the separable state $\ket{S_{1,2}}$ (cf.~Eq.~(\ref{ISMAX})), if we keep in mind the normalization factor of the state $\ket{W_{1,2}}$. Let us investigate the second sum, i.e., the quantum correlations $\langle\hat{s}^{+}_i \hat{s}^{-}_j \rangle$ of the W-state ($i \neq j$). For example, the correlation $\langle\hat{s}^{+}_1 \hat{s}^{-}_2 \rangle$ calculates to

\staffeld
\langle\hat{s}^{+}_1 \hat{s}^{-}_2 \rangle &=& \frac{1}{\sqrt 3} \left(\bra{e\,g\,g}  + \bra{g\,e\,g} + \bra{g\,g\,e} \right) \hat{s}^{+}_1 \hat{s}^{-}_2 \ket{W_{1,2}}\nonumber \\
&=&\frac{1}{3} \,\bra{g\,g\,g}\, \hat{s}^{-}_2 \left(\ket{e\,g\,g} + \ket{g\,e\,g} + \ket{g\,g\,e} \right) \nonumber\\ 
&=& \frac{1}{3} \langle g\,g\,g | g\,g\,g \rangle =  \frac{1}{3}\,.
\stoffeld

\noindent In contrast to the separable states (cf.~Eq.~(\ref{ISMAX})), the quantum correlations of the W-states are non-zero. Eq.~(\ref{IW12}) thus simplifies to

\staffeld
\label{IW12a}
\mathrm{I}_{\ket{W_{1,2}}} &=&  \frac{1}{3}(\, 3 + \sum_{\substack{i,j=1\\i \neq j}}^3 \,e^{i\,(\varphi_i-\varphi_j)}) \nonumber\\
&=& 1 + \frac{2}{3} \sum_{i<j=1}^3 \cos(\varphi_i-\varphi_j)\,, 
\stoffeld

\noindent i.e., the intensity $\mathrm{I}_{\ket{W_{1,2}}}$ displays an angular dependency and exhibits a maximum of

\staffeld
\label{IW12b}
\left[\mathrm{I}_{\ket{W_{1,2}}}\right]^{\mathrm{Max}} =  3 
\stoffeld

\noindent at $\varphi_1 = \varphi_2 = \varphi_3$, what is fulfilled for $\theta = 0\,,\,\pm \pi$ (see Eq.~(\ref{Eq3})). The maximum intensity of the W-state $\ket{W_{1,2}}$ is higher than the maximum intensity of the corresponding separable state $\ket{S_{1,0}}$ due to the fact that the quantum correlations $\langle\hat{s}^{+}_i \hat{s}^{-}_j \rangle$ are non-zero in case of the W-state. However, this is so far a rather mathematical justification and does not give much physical insight in the processes causing the enhancement. The question which we address in the next section thus is: how can we physically understand the enhancement in the emission of radiation by the entangled state compared to the separable state?

\subsection{Quantum interference inititated superradiant emission from entangled atoms}

The best way to understand the superradiant behavior from W-states is to examine the transition amplitude for each individual photon detection event. The net result would then be obtained by coherently summing over all the paths via which photons are emitted and recorded by the detector. In the following we will demonstrate that the interference of various quantum paths gives us a transparent physical picture of the superradiant emission from W-states.

\begin{figure}[h!]
\centering
\includegraphics[width=0.4 \textwidth]{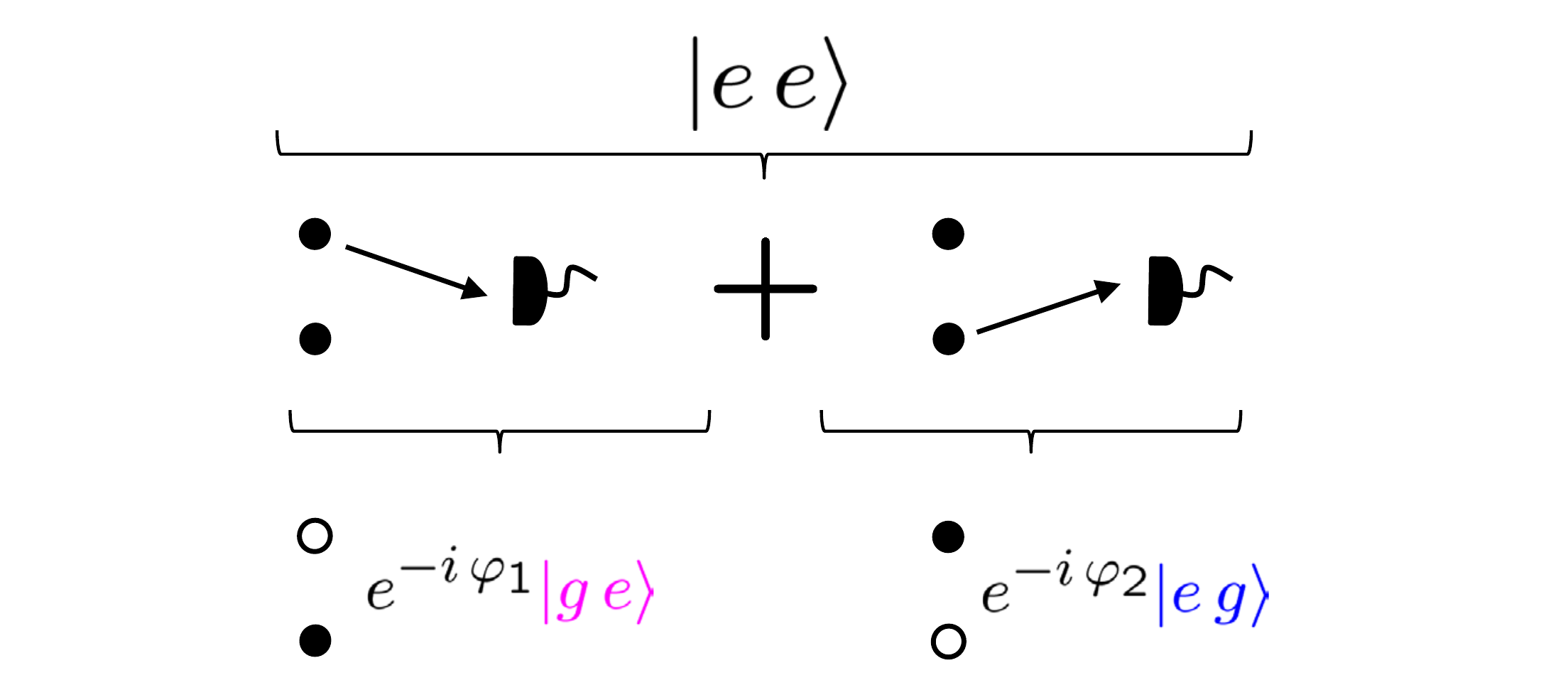}
\caption{\label{EXS} (Color online) Possible quantum paths of the initially separable state $\ket{S_{2,0}}$. Black circles denote atoms in the excited state and white circles denote atoms in the ground state. The middle row depicts the different quantum paths. The lower row displays the final states of the atoms and the phases accumulated by the photon along the different quantum paths. See text for details.}
\end{figure}

Let us first investigate the different quantum paths of the initially separable state $\ket{S_{2,0}} = \ket{e\,e}$ (cf.~Fig.~\ref{EXS}) which lead to a successful photon detection event. For a particular event the detector cannot resolve from which of the two atoms the photon was scattered due to the far-field condition. There are thus two distinct possibilities: either the photon (black arrow) was scattered by the first excited atom (black circle) transfering it into the ground state (white circle), where a phase $e^{-i\,\varphi_1}$ is accumulated by the photon, or the photon was emitted by the second atom resulting in the accumulation of the phase $e^{-i\,\varphi_2}$. Each quantum path leads to a different final state, so in principle they are distinguishable, and we do not expect interference terms in the intensity of the state $\ket{S_{2,0}}$. Explicitly, from Eq.~(\ref{ISMAX}) and Fig.~\ref{EXS}, we obtain for the intensity distribution

\staf
\label{IS20}
\mathrm{I}_{\ket{S_{2,0}}} = ||e^{-i\,\varphi_1}\color{black}\ket{g\,e}\color{black}||^2 + ||e^{-i\,\varphi_2}\color{black}\ket{e\,g}\color{black}||^2 = 2\,,
\stof

\noindent where the norm of the state vector $\ket{\Psi}$ is denoted by $||\ket{\Psi}||^2 = \langle \Psi | \Psi \rangle$. Let us compare these results to the superposition of quantum paths and the intensity distribution obtained in case of an initial W-state $\ket{W_{2,1}}$. From Eq.~(\ref{W-state}) this state reads

\staf
\ket{W_{2,1}} = \frac{1}{\sqrt3}\left(|e\,e\,g\rangle + |e\,g\,e\rangle + |g\,e\,e\rangle\right)\,.
\stof

\begin{figure}[h!]
\centering
\includegraphics[width=0.45 \textwidth]{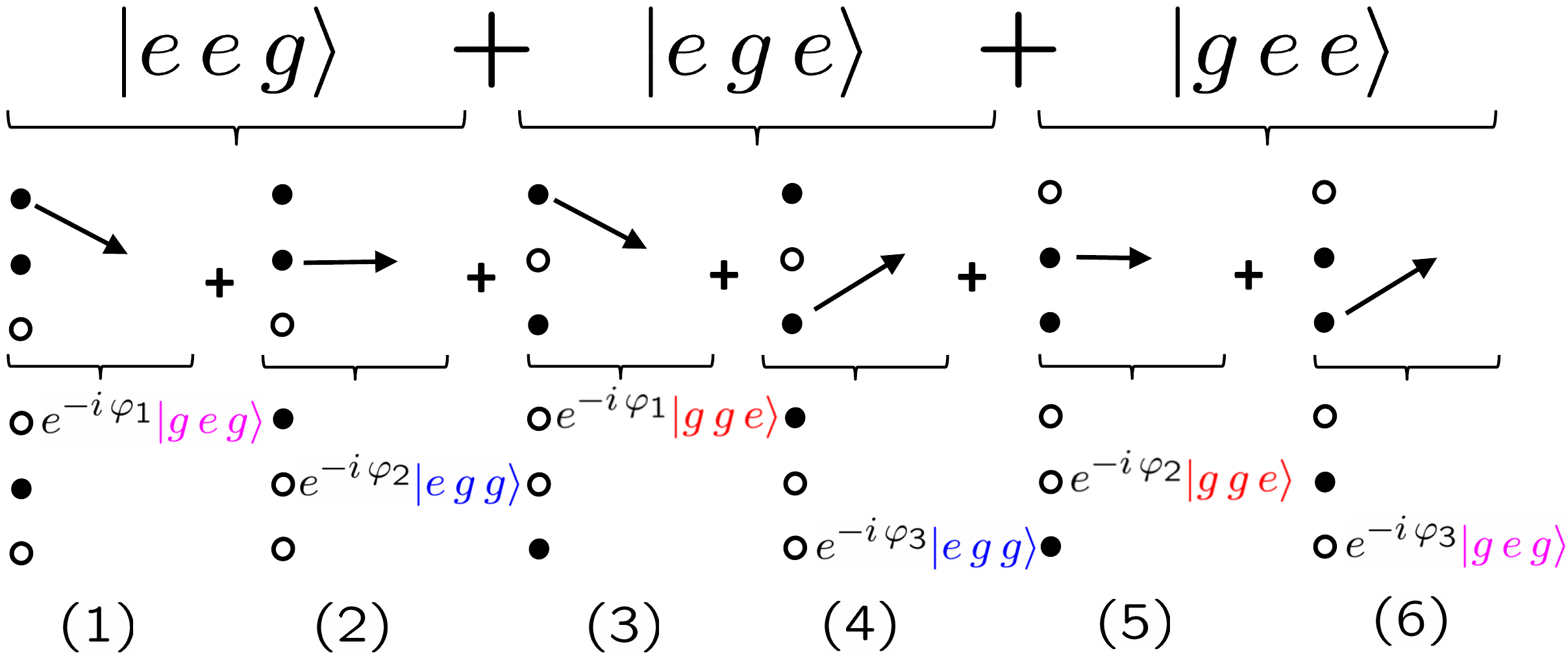}
\caption{\label{EXW} (Color online) Possible quantum paths of the initial W-state $\ket{W_{2,1}}$.}
\end{figure}

\noindent Fig.~\ref{EXW} depicts the different quantum paths leading to a successful measurement event. Let us exemplify the emerging interference by considering only the first term in the coherent sum of $\ket{W_{2,1}}$. The state $\ket{e\,e\,g}$ basically leads to the same quantum paths as the state $\ket{S_{2,0}}$: Either the first atom emits the photon leading to the final state $\ket{g\,e\,g}$ and to an accumulation of the phase $e^{-i\,\varphi_1}$ or the second atom scatters the photon, so that the final state is $\ket{e\,g\,g}$ and the accumulated phase corresponds to $e^{-i\,\varphi_2}$. However, in difference to the separable state $\ket{S_{2,0}}$ we have here a superposition of three different terms in the state $\ket{W_{2,1}}$ leading to six quantum paths in total.

\noindent These quantum paths can either lead to a constant contribution to the intensity like in the case of an initial separable state; however, they are also capable to interfere, namely with another \emph{indistinguishable quantum path}: Taking into account the normalization factor $\frac{1}{\sqrt3}$ we find the same constant contribution to the intensity in case of the W-state $\ket{W_{2,1}}$ (namely $\frac{6}{3}$ due to the six quantum paths which do not interfere) and in case of the separable state $\ket{S_{2,0}}$ (namely $2$). However, photons which were scattered from the atomic state $\ket{W_{2,1}}$ can occupy more than one quantum path leading to the same final state (cf., e.g., the far left and far right quantum path in Fig.~\ref{EXW}, both leading to the final state $\ket{g\,e\,g}$). Furthermore, for all quantum paths the initial states are equal from the detectors' point of view - due to the far-field assumption the detector is unable to identify from which atom the photon was scattered. Thus, we obtain interfering quantum paths exclusively for non-separable states which are leading to interference terms in the intensity distribution. 

Let us explicitly calculate the intensity of the state $\ket{W_{2,1}}$ to quantitatively investigate the validity of our quantum path interpretation. It reads (cf., Eq.~(\ref{IGEN}) and Fig.~\ref{EXW})

\staffeld
\label{IW21a}
\mathrm{I}_{\ket{W_{2,1}}} = \frac{1}{3}\left|\left|(e^{i\,\varphi_1}+e^{-i\,\varphi_2})\color{black}\ket{g\,g\,e}\color{black}\right|\right|^2\hspace{2.5cm}\nonumber\\
+ \frac{1}{3}\;\left|\left|(e^{-i\,\varphi_1}+e^{-i\,\varphi_3})\color{black}\ket{g\,e\,g}\color{black}\right|\right|^2 \hspace{1.5cm}\nonumber\\ + \frac{1}{3} \left|\left|(e^{-i\,\varphi_2}+e^{-i\,\varphi_3})\color{black}\ket{e\,g\,g}\color{black}\right|\right|^2\,. \hspace{0.5cm}
\stoffeld

\noindent In this section we want to focus on the maximum of the intensity distribution. From Eq.~(\ref{IW21a}) it follows

\staffeld
\label{interms}
\left[\mathrm{I}_{\ket{W_{2,1}}}\right]^{\mathrm{Max}}= \frac{6}{3}\; +\;\;\;\; \frac{6}{3}\hspace{3.5cm}\nonumber\\
\equiv \left[\mathrm{I}_{\ket{S_{2,0}}}\right]^{\mathrm{Max}} + \color{black}\mathrm{\underline{interference\,terms}}\, \color{black},
\stoffeld

\noindent in agreement with the foregoing discussion. In the following we want to demonstrate that the enhanced maximal emission of radiation scattered by W-states can be explained purely by  additional constructive interference terms created by indistinguishable quantum paths. To this end we cast the foregoing argument into a formula for the maximum of the intensity of a W-state: 

\staf
\label{genform}
\left[\mathrm{I}_{\ket{W}}\right]^{\mathrm{Max}} = \left[ \mathrm{I}_{\ket{S}}\right]^{\mathrm{Max}} + \left(\# {QP}\right) \times \left({\# \ket{f}}\right) \times  \left({\cal N}\right)\,.
\stof

\noindent Hereby, $\left({\# QP}\right)$ abbreviates the number of interfering quantum path \emph{pairs} leading to the same final state. Multiplied by the number of final states $\left(\# {\ket{f}}\right)$, we thus arrive at the total number of interfering quantum paths pairs, i.e., interference terms, contributing to the intensity maximum of the signal. Together with the squared normalization constant $\left({\cal N}\right)$ of the corresponding W-state the expression $\left(\# {QP}\right) \times \left({\# \ket{f}}\right) \times  \left({\cal N}\right)$ equals for $\theta = 0\,,\,\pm\,\pi$ the constructive contribution of the interference terms to the maximal intensity.

Let us apply Eq.~(\ref{genform}) to rederive the maximum of the intensity $\mathrm{I}_{\ket{W_{2,1}}}$. In Eq.~(\ref{IS20}) we already calculated the maximum intensity of the corresponding separable state to be $2$. The number of interfering quantum path pairs leading to the same final state can be easily obtained by counting (see Fig.~\ref{EXW}): we could either pick out the two pairs $(1,6)$ and $(6,1)$ \emph{or} the pairs $(2,4)$ and $(4,2)$ \emph{or} the two pairs $(3,5)$ and $(5,3)$, i.e., $\left(\# {QP}\right) = 2$. Note that the pairs $(i,j)$ equal to interfering quantum paths $e^{i(\varphi_i - \varphi_j)}$ for $i \neq j$ giving rise to interference terms. The number of different final states is $\left(\# {\ket{f}}\right) = 3$ and the squared normalization of the state $\ket{W_{2,1}}$ is $\left({\cal N}\right) = \frac{1}{3}$. Thus, we obtain for the maximum intensity (cf.~Eq.~(\ref{IW21a}))

\staf
\label{IW21b}
\left[\mathrm{I}_{\ket{W_{2,1}}}\right]^{\mathrm{Max}} = 2 + 2 \times 3 \times  \frac{1}{3} = 4\,,
\stof

\noindent in accordance with Eq.~(\ref{interms}).


Now we adopt the foregoing reasoning to an initial generalized symmetric W-state $\ket{W_{n_e,N-n_e}}$ with $n_e$ excited atoms and $N-n_e$ atoms in the ground state (cf., Eq.~(\ref{W-state})). The general formula for the maximum intensity of the W-state $\ket{W_{n_e,N-n_e}} \equiv \ket{W_{\star}}$ can be derived using combinatorial considerations and the maximum of the intensity of the separable state $\mathrm{I}_{\ket{S_{n_e,N-n_e}}}$ as given in Eq.~(\ref{ISMAX}). It reads

\staffeld
\label{IWMAX1}
\left[\mathrm{I}_{\ket{W_{\star}}}\right]^{\mathrm{Max}} &=& n_e  + \left[\left(\# {QP}\right) \times \left({\# \ket{f}}\right) \times  \left({\cal N}\right)\right]_{\ket{W_{\star}}} \nonumber\\
&=& n_e + \color{black}n_g(n_g+1) \times \color{black}{N \choose n_e-1} \times {N \choose n_e}^{-1} \nonumber\\
&=& n_e (n_g + 1)\,.\hspace{3cm}
\stoffeld

\noindent Let us investigate the different terms of Eq.~(\ref{IWMAX1}) in more detail. As stated before $\left({\cal N}\right)$ is the squared normalization constant of the generalized symmetric W-state (cf., Eq.~(\ref{W-state})). The number of final states $\left({\# \ket{f}}\right)$ can be derived by taking into account that after the detection of a photon there are $n_e -1$ excited atoms left which are able to occupy $N$ different position in the chain of $N$ atoms, what leads to ${N \choose n_e - 1}$. The crucial term $\left(\# {QP}\right)$ needs more explanation: $n_e$ different single quantum paths are leading to a detection event for every term of the initial W-state (cf., e.g., Eq.~(\ref{IW21a})). If we now multiply these single quantum paths by the number of terms of the initial W-state (given by ${N \choose n_e}$) we arrive at the total number of single quantum paths. The number of single quantum paths leading to the same final states - abbreviated by $\left(\# {sqp}\right)$ - is then obtained if we divide the total number of single quantum paths by the number of final states: 

\staf
\left(\# {sqp}\right) = \frac{n_e \, {N \choose n_e}}{ {N \choose n_e-1}} = n_g + 1\,.
\stof

\noindent These $n_g + 1$ single quantum paths which lead to the same final state now interfere among each other producing in total $\left(\# {QP}\right) = n_g(n_g + 1)$ interfering quantum path pairs. Clearly $\left(\# {QP}\right)$, i.e., the enhancement of the intensity is zero, if $n_g =0$. Table \ref{QPoverview} displays the results of Eq.~(\ref{IWMAX1}) for $n_e = 1,...,N-1$. 

\begin{table}[h!]
\centering
\includegraphics[width=0.4 \textwidth]{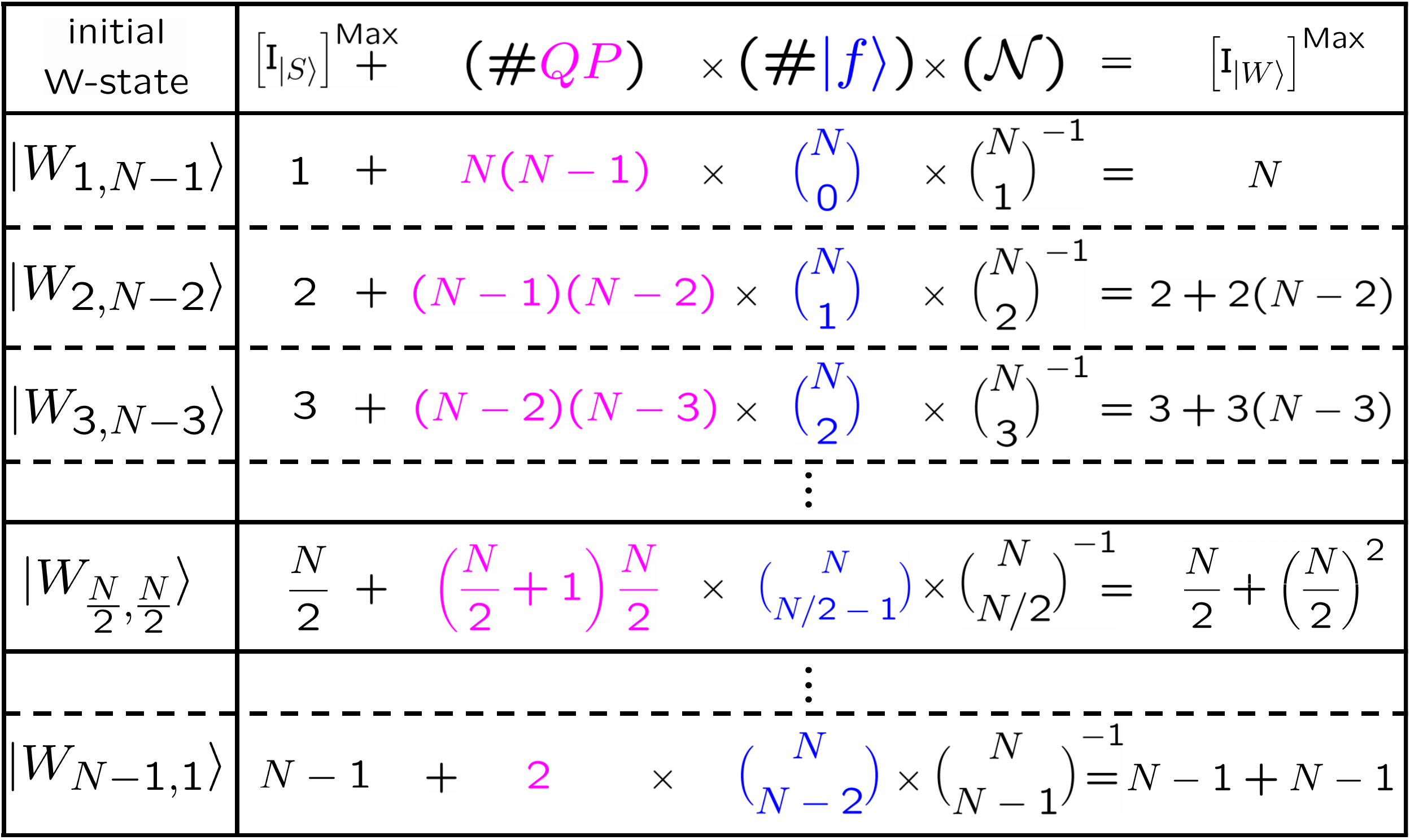}
\caption{\label{QPoverview} (Color online) Maximum intensity for the W-states $\ket{W_{n_e,N-n_e}}$ with $n_e = 1,...,N-1$. See text for details.}
\end{table}

\begin{figure}[h!]
\centering
\includegraphics[width=0.4 \textwidth]{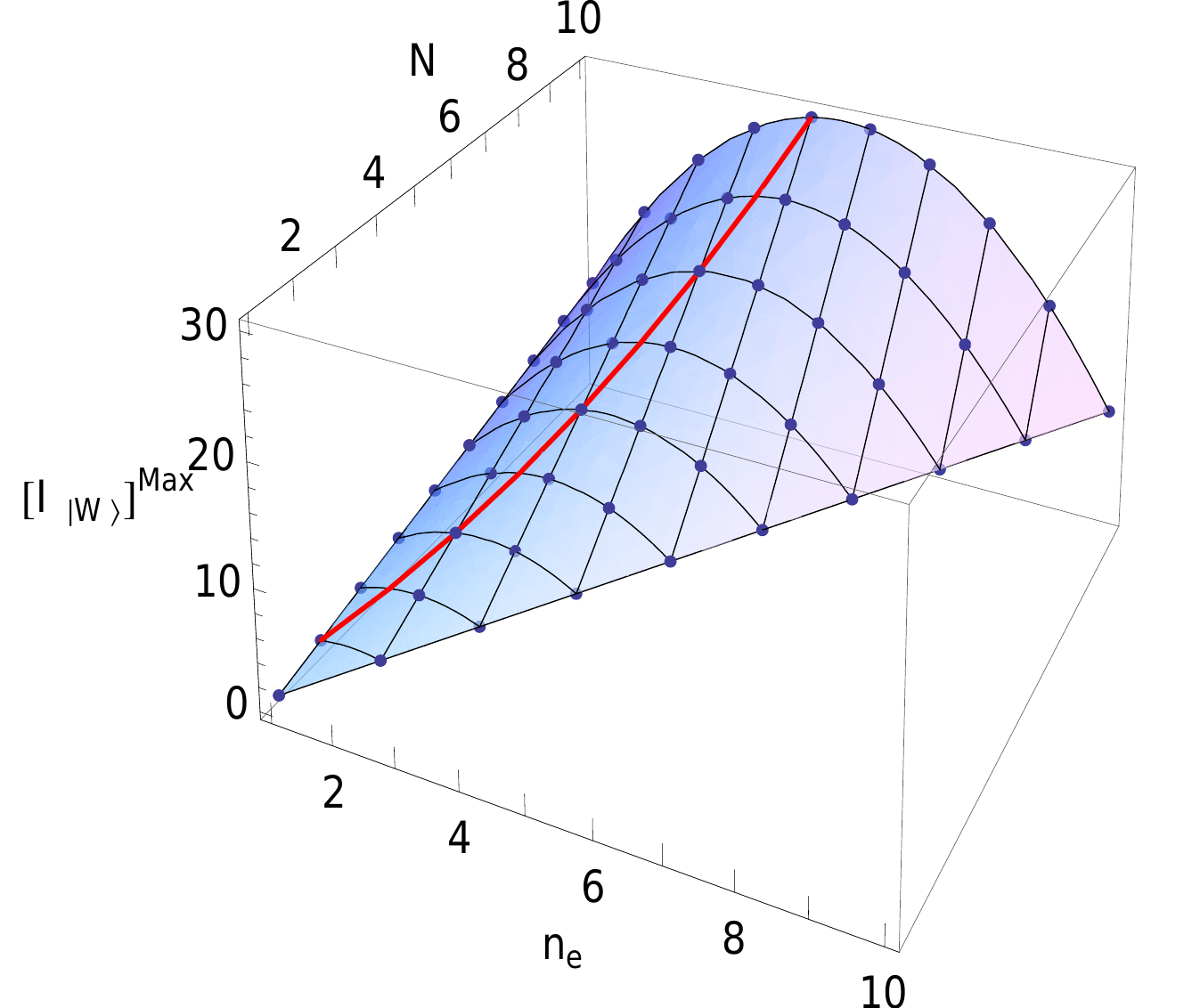}
\caption{\label{PLOTMAX} (Color online) Pointplot of the maximum of $\mathrm{I}_{\ket{W_{n_e,N-n_e}}}$ as a function of $n_e$ and $N = n_e + n_g$ (the surface serves only to guide the eye). $\left[\mathrm{I}_{\ket{W_{N/2,N/2}}}\right]^{\mathrm{Max}}$, given in units of the intensity produced by a one atom system, is highlighted by the solid line.}
\end{figure}

\noindent Furthermore, a 3d-plot of the maximum intensity of the generalized W-state $\ket{W_{n_e,N-n_e}}$ as a function of $n_e$ and $N$ (cf., Eq.~(\ref{IWMAX1})) is shown in Fig.~\ref{PLOTMAX}, where the maximum intensity of the state $\ket{W_{N/2,N/2}}$ is highlighted by the solid line.

\noindent Let us next define an enhancement parameter $\epsilon$ which describes the ratio of the maximum intensity of a W-state and the maximum intensity of a separable state with the same number $n_e$ of initially excited atoms. With Eq.~(\ref{ISMAX}) and (\ref{IWMAX1}) it calculates to $\epsilon = n_g + 1$, what is clearly greater than $1$ for $n_g \neq 1$, i.e., every W-state radiates stronger than the corresponding separable state. This behavior seems quite counterintuitive: the addition of an unexcited atom to the fully excited $N-1$ qubit compound increases the emission of the system by a factor of $2$ \cite{Dicke54} - however, just as long as there is in principle no information available about what particular atom is unexcited.

\begin{table}[h!]
\centering
\includegraphics[width=0.4 \textwidth]{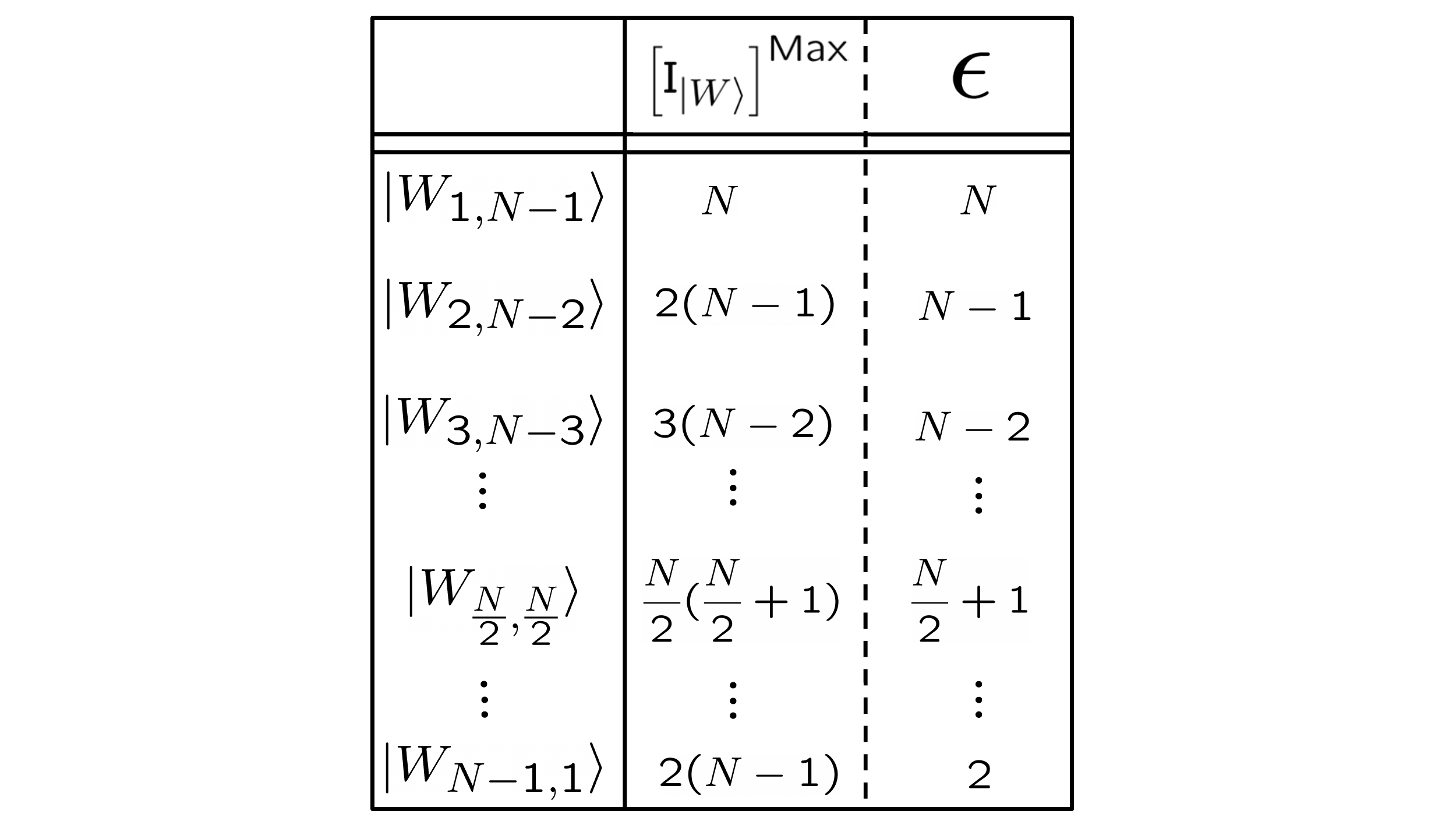}
\caption{\label{enoverview} Maximal intensity $\left[\mathrm{I}_{\ket{W}}\right]^{\mathrm{Max}}$ and enhancement parameter $\epsilon$ of the states $ \ket{W_{n_e,N-n_e}}$ with $n_e = 1,\ldots,N-1$.}
\end{table}

Note that the highest enhancement $\epsilon$ is achieved with the W-state $\ket{W_{1,N-1}}$ (cf.~Table \ref{enoverview}). In fact, the enhancement $\epsilon$ is of the order of $N$ for all generalized symmetric W-states. However, the maximum intensity for an even number of atoms, namely $N/2(N/2 +1)$, is obtained with the initial atomic states $\ket{W_{N/2,N/2}}$ (cf., the solid line in Fig.~\ref{PLOTMAX}) or $\ket{W_{N/2 + 1,N/2 - 1}}$. For an odd number of atoms the $N$-qubit state $\ket{W_{(N+1)/2,(N-1)/2}}$ even exhibits an intensity maximum of $(\frac{N + 1}{2})^2$. 

\subsection{Quantum multipath interference and Dicke superradiance}
\label{DickeSEC}

We now establish the connection between our work and that of Dicke \cite{Dicke54}. Dicke used the addition of angular momentum algebra to introduce the collective states $\ket{S,m,\nu}$ for which the collective spin operators $\hat{{\bf S}}^2$ and their $z$ component $\hat{{\bf S}}_z$ have eigenvalues $\hbar^2 s(s+1)$ and $\hbar m$, respectively, and $\nu$ is a degeneracy parameter \cite{MandelWolf95}. For $S = N/2$, where $N$ corresponds to the number of two-level atoms in the system, the state is fully symmetric and non-degenerate. These states are in fact identical to the generalized symmetric W-states $\ket{W_{n_e,N-n_e}}$ introduced above, where in our notation $m = \frac{1}{2}(n_e - n_g) = n_e - \frac{N}{2}$. Dicke assumed that the size of the system is much smaller than a wavelength and showed that in this case the radiation rate from the state $\ket{N/2,m}$ is

\staf
\label{DickeEQ}
\mathrm{I} \propto (\frac{N}{2} + m)(\frac{N}{2} - m + 1)\,.
\stof

\noindent For $m = 0$ the radiation rate is clearly of the order of $N^2$. In the previous section we have found a similar result for the symmetric W-states, in particular for the state $\ket{W_{N/2,N/2}}$ corresponding in Dicke's notation to the state $|S = N/2, m=0,\nu =0\rangle$. However we specifically do not consider the limit of small systems as then the dipole-dipole interaction between the atoms is to be accounted for and this completely changes the radiation properties \cite{Friedberg73}. We rather consider the case where the system size is larger than a wavelength to avoid the difficulties due to the dipole-dipole interaction.

Usually an enhanced radation rate is related to a large dipole moment. Contrary to this, the Dicke states have zero dipole moment, but show a superradiant behavior. So far, a clear physical understanding of this remarkable result is missing. Our physical picture based on quantum multipath interference is able to explain the superradiant behavior: The entanglement in the W-states leads to constructive interference between different indistinguishable pathsways which then is responsible for the superradiant emission. We can explicitly write down all the interfering pathways for any number of atoms and for any initial state. We stress that the situation that we discuss is different from the way earlier experiments have been performed. In common experiments on superradiance \cite{SUPEXPS} a gas of atoms is initially prepared in the Dicke state $\ket{N/2,N/2}$ (the fully excited separable state $\ket{S_{N,0}}$ in our notation) in order to investigate the variation of the systems' radiation over a long time scale as the state evolves to the ground state. In contrast, we are investigating the short time behavior of the radiation emitted by well prepared initial W-states.

\section{Directionality in the emission from arbitrary symmetric W-states}
\label{direct}

So far we focussed our discussion on establishing a physical explanation for the maxima of the intensity radiated by generalized symmetric W-states. However, besides studying the maximal enhancement of radiation we now want to investigate the angular dependence of the scattered intensity to better characterize the radiation emitted by those states. Using Eq.~(\ref{IGEN}) the intensity distribution of a generalized symmetric W-state calculates to (cf.~Appendix \ref{app1})

\staffeld
\label{IWPHI}
\mathrm{I}_{|W_{n_e,N-n_e}\rangle}(\theta) = \frac{n_e\,(n_e - 1)}{N - 1} + \frac{n_e\,(N - n_e)}{N(N-1)}  \frac{\sin^2(\frac{\varphi_N}{2})}{\sin^2(\frac{\varphi_1}{2})}\,.\nonumber\\ \nonumber\\
\stoffeld

\noindent With $\delta \varphi_N = N\,k\,d\,\cos{\theta}\, \delta\theta$ (cf.~Eq.~(\ref{Eq3})) the width of the interference term in Eq.~(\ref{IWPHI}) is given by

\staf
\label{width}
\delta \theta = \frac{2\pi}{N\,k\,d}\,.
\stof

\noindent The width of the distribution thus dependens on the wavelength $\lambda$, the distance $d$ between adjacent atoms and on the total number of atoms $N$ - in contrast to the enhancement of the emitted intensity which depends on the product of unexcited atoms $n_g$ and excited atoms $n_e$ (cf.~Eq.~(\ref{IWMAX1})).

Let us introduce the visibility of the intensity distribution 

\staf
{\cal V} = \frac{\mathrm{I}_{\mathrm{Max}} - \mathrm{I}_{\mathrm{Min}}}{\mathrm{I}_{\mathrm{Max}} + \mathrm{I}_{\mathrm{Min}}}\,.
\stof

\noindent From Eq.~(\ref{IWPHI}) we find

\staffeld
\label{VISW}
{\cal V} = \frac{1}{1 + \frac{2(n_e-1)}{N\,n_g}}\,.
\stoffeld

\noindent For $n_e = 1$ the visibility is $1$. Furthermore, for a given number of excited atoms $n_e$ and an increasing number of unexcited atoms $n_g$ sharing a common W-state ${\cal V} \rightarrow 1$. This behavior can be understood from Eq.~(\ref{genform}) as the offset of the intensity distribution only depends on $n_e$.

Let us investigate the case $n_e = 1$ in more detail. Eq.~(\ref{IWPHI}) becomes in this case proportional to the intensity distribution of a diffraction grating 

\staf
\mathrm{I}_{|W_{1,N-1}\rangle}(\theta) =  \frac{1}{N}  \frac{\sin^2(\frac{\varphi_N}{2})}{\sin^2(\frac{\varphi_1}{2})}\,.
\stof


\begin{figure}[h!]
\centering
\includegraphics[width=0.45 \textwidth]{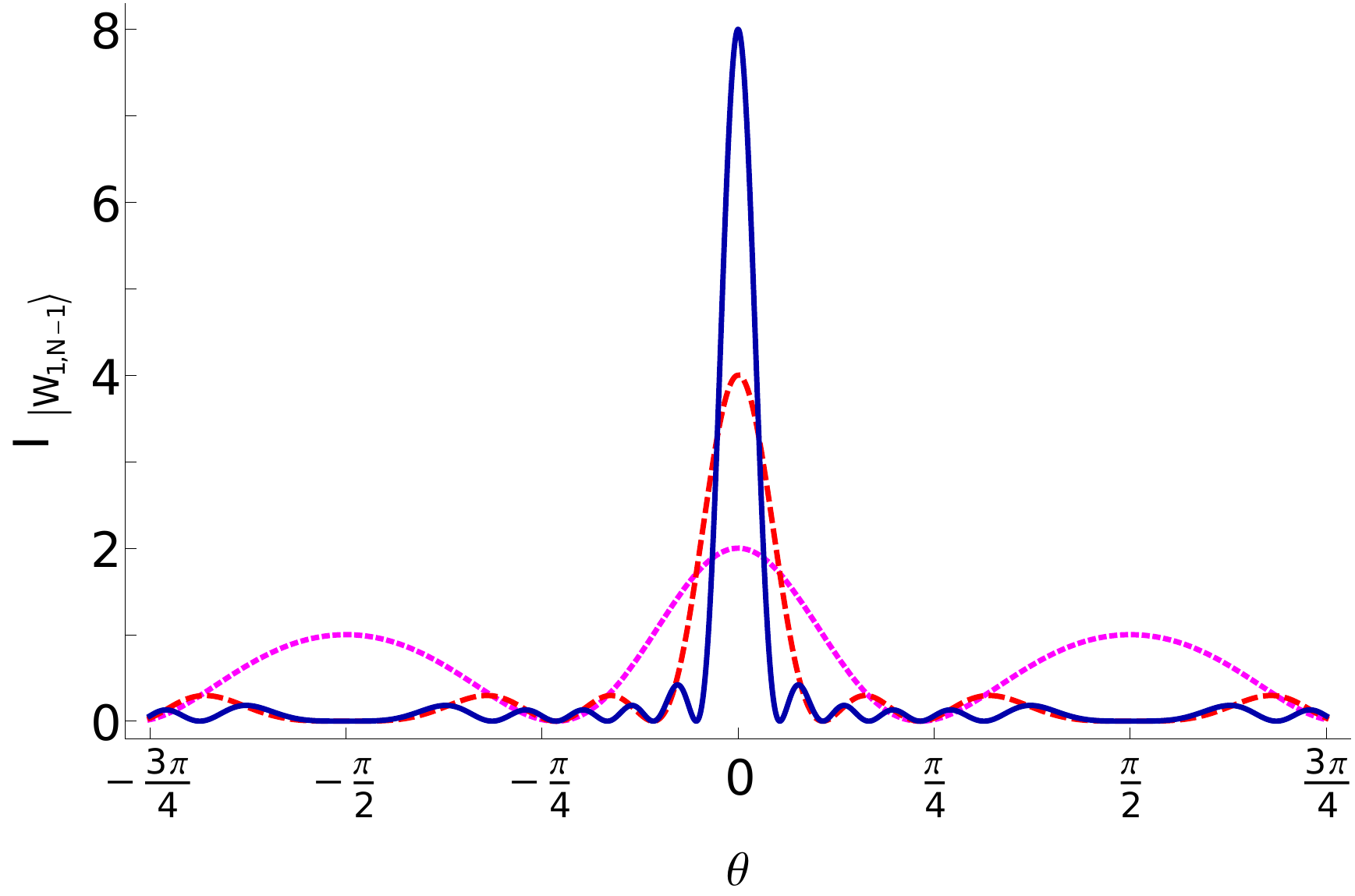}
\caption{\label{PLOTDIR1} (Color online) Intensity distribution $\mathrm{I}_{|W_{1,N-1}\rangle}(\theta)$ of the state $|W_{1,N-1}\rangle$ as a function of the detector position $\theta$ (cf.~Fig.~\ref{setup}) for different numbers of atoms $N$ ($N = 2$ dotted, $N = 4$ striped, $N = 8$ solid) and with $k\,d = \frac{3}{2}\pi$. $\mathrm{I}_{|W_{1,N-1}\rangle}$ is given in units of the intensity produced by a one atom system. A strong directionality in the emission of radiation by the symmetric W-states can be seen.}
\end{figure}

\noindent Fig.~\ref{PLOTDIR1} displays $\mathrm{I}_{|W_{1,N-1}\rangle}$ for different numbers of atoms $N = n_e + n_g$. The figure clearly shows that with a higher number of unexcited atoms $n_g = N -1$ the radiation of the state $|W_{1,N-1}\rangle$ is increasingly peaked in the directions $\theta = 0$ (see Eq.~(\ref{IWMAX1})). Furthermore, we can identify certain directions where the intensity vanishes completely (leading to a visibility of $1$) and large areas where the intensity is almost zero. In comparison to the superradiant peak at $\theta = 0$ these areas are subradiant. Thus, we obtain also with symmetric generalized W-states a subradiant behavior in certain directions, what is quite counterintuitive as this phenomenon is in general exclusively ascribed to anti-symmetric states, i.e., to Dicke states $\ket{S,m}$ with $S \neq N/2$. 

\begin{figure}[h!]
\centering
\includegraphics[width=0.45 \textwidth]{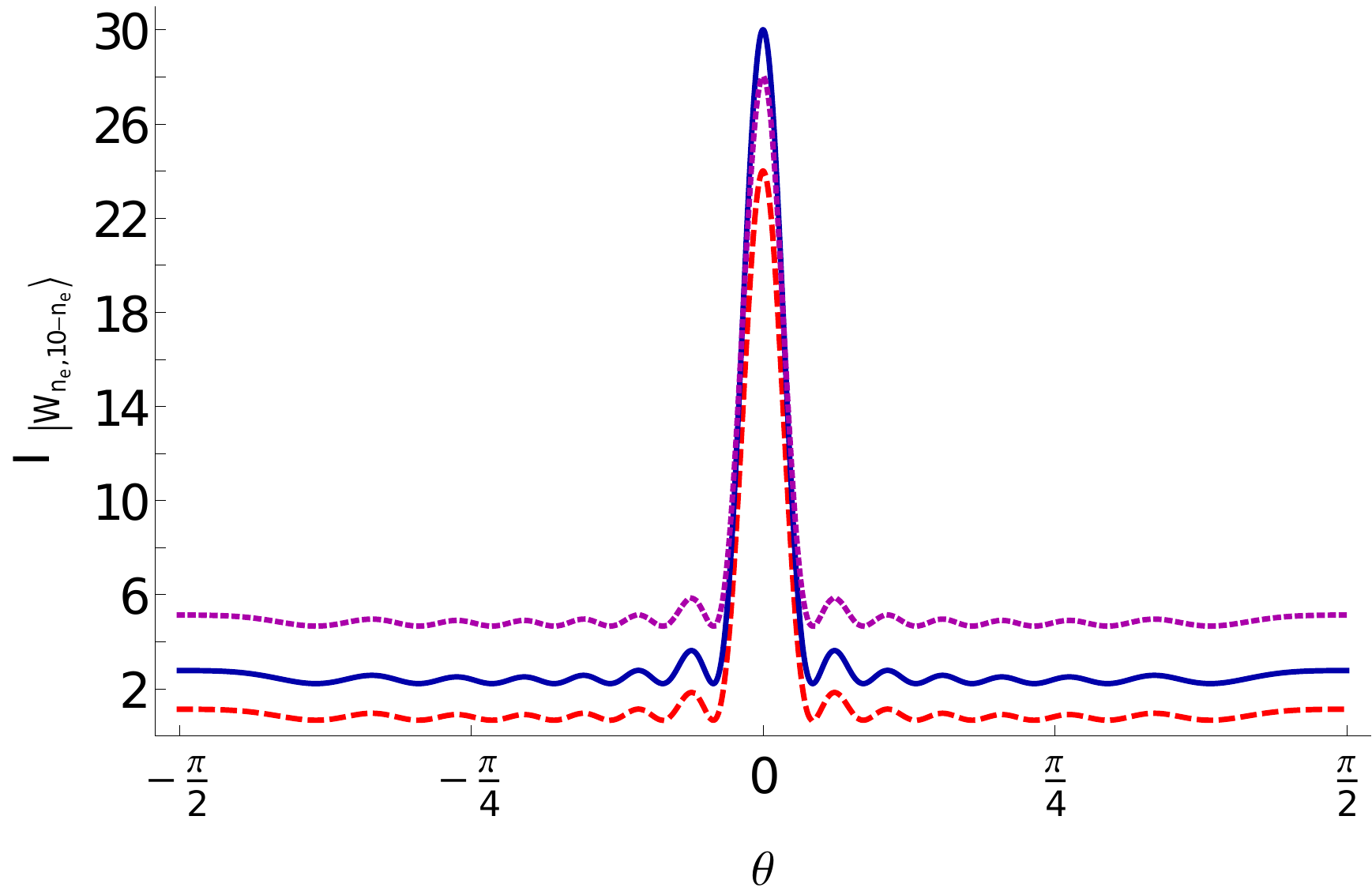}
\caption{\label{PLOTDIR2} (Color online) Intensity distribution of the state $|W_{n_e,10-n_e}\rangle$ for $N = 10$ and different numbers of excited atoms $n_e$ ($n_e = 3$ striped, $n_e = 5$ solid, $n_e = 7$ dotted); $k\,d = \frac{3}{2}\pi$ and $\mathrm{I}_{|W_{n_e,10-n_e}\rangle}$ is given in units of the intensity produced by a one atom system.}
\end{figure}

Let us discuss a second example, namely the intensity distribution of an initial atomic state $|W_{n_e,N-n_e}\rangle$ for a fixed number of atoms $N$ and varying number of excited atoms $n_e$ (see Fig.~\ref{PLOTDIR2}). For $N = 10$, $n_e = 5$ and $n_g = 5$, i.e., the initial state $|W_{N/2,N/2}\rangle$ (solid line), we find at $\theta = 0$ the maximum of all graphs as expected (cf.~Eq.~(\ref{IWMAX1})). Furthermore, it can be seen that the width of the distributions is independent of $n_e$ and equal for all graphs since it just depends on the total number of atoms $N$ sharing the W-state (cf.~Eq.~(\ref{width})). Finally, as expected, the visibility is decreasing with increasing number of excited atoms $n_e$ (cf., Eq.~(\ref{VISW})). 

\begin{figure}[htbp]
  \centering
  \subfigure{
    \label{w14c1}
    \includegraphics[width=0.45 \textwidth]{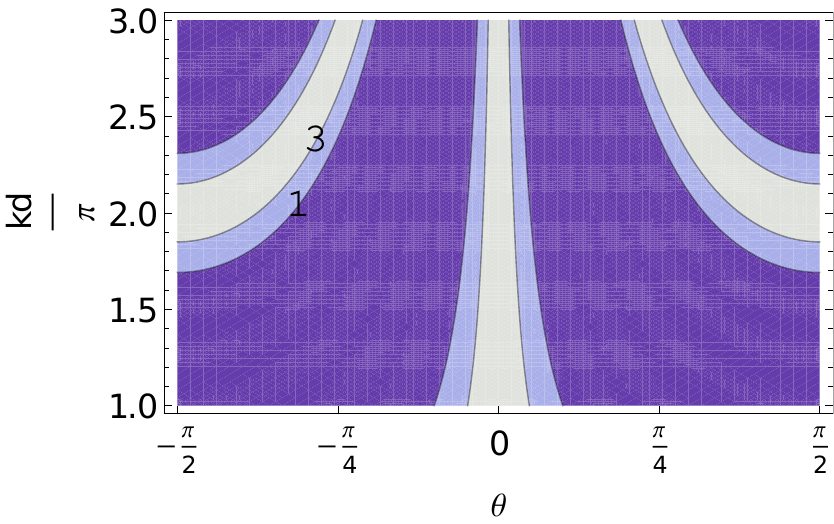}
  }\qquad
  \subfigure{
    \label{w14c2}
    \includegraphics[width=0.45 \textwidth]{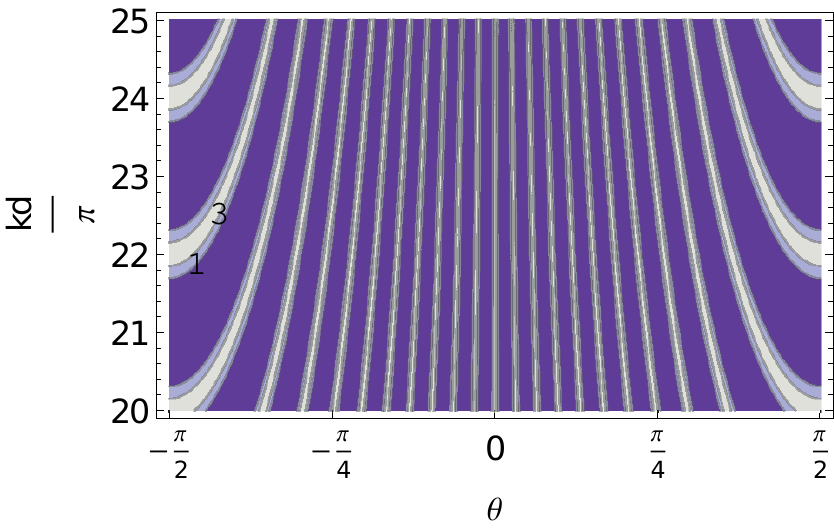} 
  }
  \caption{(Color online) Contourplots of the intensity distribution $\mathrm{I}_{|W_{1,4}\rangle}$ of the state $|W_{1,4}\rangle$ dependent on the observation angle $\theta$ and the interatomic spacing $k\,d$ in units of $\pi$. The numbers $1$ and $3$ on the left hand side in each plot indicate the two contours $\mathrm{I}_{|W_{1,4}\rangle} = 1$ and $\mathrm{I}_{|W_{1,4}\rangle} = 3$, i.e., dark areas correspond to a low intensity, bright areas to a high intensity. $\mathrm{I}_{|W_{1,4}\rangle}$ is given in units of the intensity produced by a one atom system.}
  \label{c1c2}
\end{figure}

We want to conclude this section by discussing a variation of the interatomic spacing $d$. In the upper part of Fig.~\ref{c1c2} a contourplot of the intensity distribution $\mathrm{I}_{|W_{1,4}\rangle}$ of the state $|W_{1,4}\rangle$ is shown as a function of the observation angle $\theta$ and the interatomic spacing $k\,d$ (cf., the striped line in Fig.~\ref{PLOTDIR1}). The superradiant maximum can be clearly seen at $\theta = 0$. However, there are also periodically appearing sub- and superradiant areas at $\theta \approx \frac{\pi}{2}$ which lead to an increasing number of fringes with increasing $k\,d$. Under experimental conditions typical distances between the atoms are of the order of $10 \lambda$ \cite{BlattWineland08}, what leads to $k\,d \approx 20\,\pi$. A corresponding contourplot of the intensity distribution $\mathrm{I}_{|W_{1,4}\rangle}$ is plotted from $k\,d = 20\pi$ to $k\,d = 25\pi$ in the lower part of Fig.~\ref{c1c2}. As expected, the fringe spacing decreases with increasing atom separation - however, the particularities of the angular dependence remain unchanged, even if we consider larger distances between the atoms.

Finally we note that in the limit of large $N$, the function $\frac{\sin^2(\frac{\varphi_N}{2})}{\sin^2(\frac{\varphi_1}{2})}$ (cf., Eq.~(\ref{IWPHI})) becomes sharply peaked at $\varphi_1 = 0$ or $\theta = 0$. For a periodic system, where the assumption of large $N$ is implicit, a similar result is obtained in \cite{Nienhuis87}. Further, for an ensemble of atoms angular distributions have been extensively studied \cite{SUPEXPS,Scully06,Scully09}.



\section{Radiation from anti-symmetric W-states - competition of destructive and constructive interference}
\label{antiW}

In this section we want to apply our approach of using quantum interferences for the derivation of the enhanced intensity generated by entangled states also to anti-symmetric W-states, i.e., Dicke states $\ket{S,m,\nu}$ with $S \neq N/2$. As there is no analytical expression for these states we cannot present a general formula for the different contributions $\left(\# {QP}\right)$, $\left(\# {\ket{f}}\right)$ and $\left({\cal N}\right)$ like in Eq.~(\ref{IWMAX1}). However, we can motivate that our physical framework also gives good predictions in case of anti-symmetric states by investigating an explicit example: let us consider the chain of atoms to be initially in the anti-symmetric state

\staf
\ket{W^-_{2,1}} = \frac{1}{\sqrt6}(\left|e\,g\,e\rangle + |e\,e\,g\rangle - 2\,|g\,e\,e\rangle\right)\,.
\stof

\noindent From Eq.~(\ref{IGEN}), the intensity distribution of this state calculates to

\staffeld
\label{IANTIW}
\mathrm{I}_{\ket{W^-_{2,1}}} = \frac{1}{6}\left|\left|(e^{i\,\varphi_1}-2\,e^{i\,\varphi_2})\color{black}\ket{g\,g\,e}\color{black}\right|\right|^2\hspace{1cm } \nonumber\\ 
+\frac{1}{6}\left|\left|(e^{i\,\varphi_1}-2\,e^{i\,\varphi_3})\color{black}\ket{g\,e\,g}\color{black}\right|\right|^2 \hspace{0.5cm }\nonumber\\ +\frac{1}{6}\left|\left|(e^{i\,\varphi_2}+e^{i\,\varphi_3})\color{black}\ket{e\,g\,g}\color{black}\right|\right|^2 \nonumber\\
= \frac{1}{6}\left\{ 12 - 4 \left[ \cos(\color{black}\varphi_1 - \varphi_2\color{black}) + \cos(\color{black}\varphi_1 - \varphi_3\color{black})\right] \right.\nonumber\\
+\left. 2\cos(\color{black}\varphi_2 - \varphi_3\color{black})\right\}\,.
\stoffeld

\noindent Since we want to investigate the subradiant behavior of this state, we are now looking for the minimum of Eq.~(\ref{IANTIW}). It is found to be 

\begin{figure}[h!]
\centering
\includegraphics[width=0.47 \textwidth]{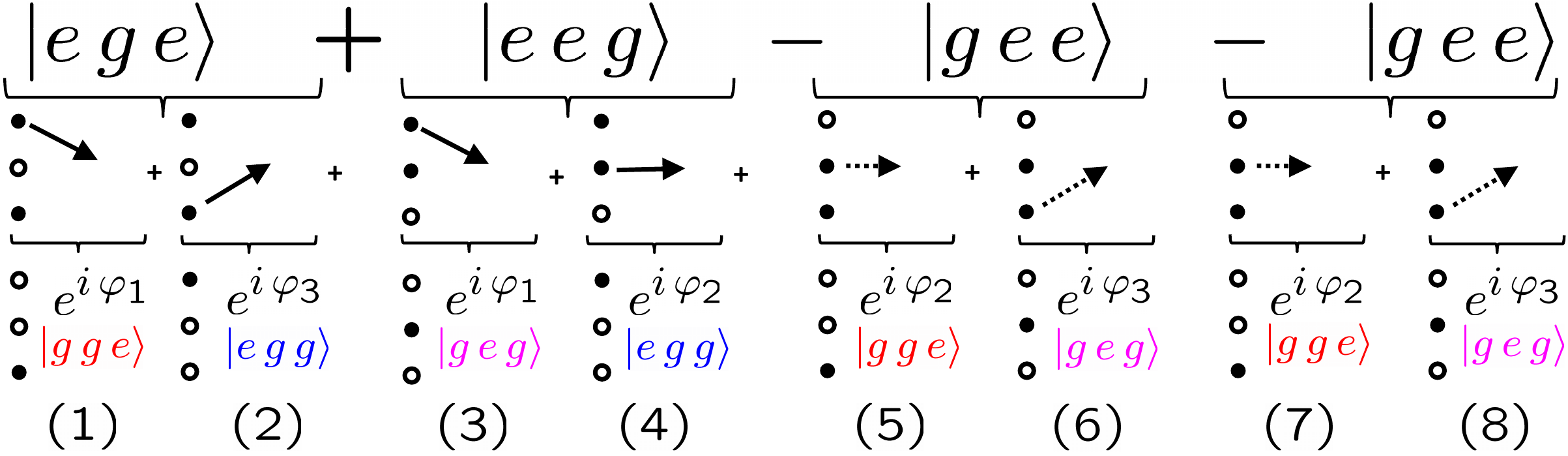}
\caption{\label{EXAW} (Color online) Quantum paths of the anti-symmetric state $\ket{W^-_{2,1}}$. Black circles denote atoms in the excited state and white circles denote atoms in the ground state. The middle row depicts the different quantum paths. The lower row displays the final states of the atoms and the phases accumulated by the photon due to different quantum paths. Quantum paths leading to constructive interference are denoted by solid arrows and quantum paths leading to destructive interference are depicted by dashed arrows. See text for details.}
\end{figure}

\staf
\label{IANTIW1}
\left[\mathrm{I}_{\ket{W^-_{2,1}}}\right]^{\mathrm{Min}} = 1\,,
\stof

\noindent at the detector positions $\theta = 0\,,\,\pm\,\pi$. By examining Eq.~(\ref{IANTIW}) and by studying the corresponding quantum paths contributing to $\mathrm{I}_{\ket{W^-_{2,1}}}$ (see Fig.~\ref{EXAW}), it can be seen that we again find the same offset in case of the anti-symmetric W-state $\ket{W^-_{2,1}}$ (namely $\frac{12}{6}$) as in case of the separable state $\ket{S_{2,0}}$ or the corresponding symmetric W-state $\ket{W_{2,1}}$ (cf.~section \ref{enhancedW}). However, in difference to the symmetric states, where all quantum paths led to constructive interference at $\theta = 0\,,\,\pm\,\pi$, we now have two different types of interfering quantum paths: due to the different signs of the prefactors in Eq.~(\ref{IANTIW}) some paths are leading to constructive interference, other quantum paths to destructive interference (denoted by solid and dashed lines in Fig.~\ref{EXAW}).

Incorporating this particularity we now show that the reduced emission of radiation of anti-symmetric W-states can again be explained by the interference of indistinguishable quantum paths. For this purpose we conjecture in analogy to Eq.~(\ref{genform}) that the radiation from an asymmetric W-state can be expressed as

\staffeld
\label{genform1}
\left[\mathrm{I}_{\ket{W^-}}\right]^{\mathrm{Min}} = \left[ \mathrm{I}_{\ket{S}}\right]^{\mathrm{Max}} + \left\{\left(\# {QP}_C\right) \times \left({\# \ket{f}}_C\right)\right. \nonumber\\ 
- \left.\left(\# {QP_D}\right) \times \left({\# \ket{f}}_D\right)\right\} \times  \left({\cal N}\right)\,,
\stoffeld

\noindent where $\left({\# QP}_C\right)$ abbreviates the number of constructive interfering quantum path pairs leading to \emph{one} of the possible final states. Further, $\left({\# QP}_D\right)$ denotes the number of destructive interference terms leading to one final state. Multiplied by $\left(\# {\ket{f}}_C\right)$ or by $\left(\# {\ket{f}}_D\right)$, which denotes the number of corresponding final states, we thus arrive at the contribution of constructive and destructive interfering quantum path pairs to the intensity of the signal. Again, $\left({\cal N}\right)$ is the squared normalization constant of the corresponding anti-symmetric W-state.

Let us apply Eq.~(\ref{genform1}) to the example discussed above to obtain the minimum of the intensity $\mathrm{I}_{\ket{W^-_{2,1}}}$. The maximum intensity of the corresponding separable state is $\mathrm{I}_{\ket{S_{2,0}}} = 2$ (cf.~Eq.~(\ref{IS20})). The number of constructive and destructive interfering quantum path pairs leading to their corresponding final states can be again obtained by counting (see Fig.~\ref{EXAW}): we count in case of constructive interference (only solid arrows) the two pairs $(2,4)$ and $(4,2)$, i.e., $\left(\# {QP}_C\right) = 2$ and $\left(\# {\ket{f}}_C\right) = 1$. The destructive interfering quantum path pairs (solid and dashed arrows) would be $(1,5)$, $(5,1)$, $(1,7)$ and $(7,1)$ \emph{or} the pairs $(3,6)$, $(6,3)$, $(3,8)$ and $(8,3)$, i.e., $\left(\# {QP}_D\right) = 4$ and $\left(\# {\ket{f}}_D\right) = 2$. With the squared normalization $\left({\cal N}\right) = \frac{1}{6}$ of the state $\ket{W^-_{2,1}}$ we thus again obtain for the minimum intensity (cf.~Eq.~(\ref{IANTIW1}))

\staf
\label{IANTIW2}
\left[\mathrm{I}_{\ket{W^-_{2,1}}}\right]^{\mathrm{Min}} = 2  + (2\times1 - 4 \times 2) \times \frac{1}{6} = 1\,.
\stof

Let us conclude this section by comparing the angular intensity distribution of the state $\ket{W^-_{2,1}}$, the state $\ket{W_{2,1}}$ and the state $\ket{\tilde{W}^{-}_{{2,1}}}$, where the latter is given by

\staf
\ket{\tilde{W}_{2,1}} = \frac{1}{\sqrt2} \ket{e} \otimes \left( \ket{g\,e} - \ket{e\,g}\right)\,. 
\stof

\noindent The state $\ket{\tilde{W}^{-}_{2,1}}$ shows the same intensity minimum as the anti-symmetric state $\ket{W^-_{2,1}}$: In Dicke notation the $S$ and the $m$ parameter of the two states are equal, namely $S=1/2$ and $m = 1/2$, leading to a minimum intensity of \cite{Dicke54}

\staf
\label{dickesub}
I \propto (S + m) (S - m + 1) = 1\;.
\stof
\noindent However, the angular intensity distribution of the state $\ket{\tilde{W}^{-}_{2,1}}$ differs from the one of the state $\ket{{W}^{-}_{2,1}}$ (cf., Eq.~(\ref{IANTIW})) and calculates to

\staffeld
\label{IANTIW3}
\mathrm{I}_{\ket{\tilde{W}^{-}_{2,1}}} = \frac{1}{2}\left|\left|(e^{-i\,\varphi_1}(\color{black}\ket{g\,g\,e} - \ket{g\,e\,g}\color{black})\right|\right|^2\hspace{0.5cm}\nonumber\\ +\frac{1}{2}\left|\left|(e^{-i\,\varphi_3}-\,e^{-i\,\varphi_2})\color{black}\ket{e\,g\,g}\color{black}\right|\right|^2 \nonumber \\ 
= 2 - \cos(\color{black}\varphi_2 - \varphi_3\color{black})\,.
\stoffeld

\begin{figure}[h!]
\centering
\includegraphics[width=0.45 \textwidth]{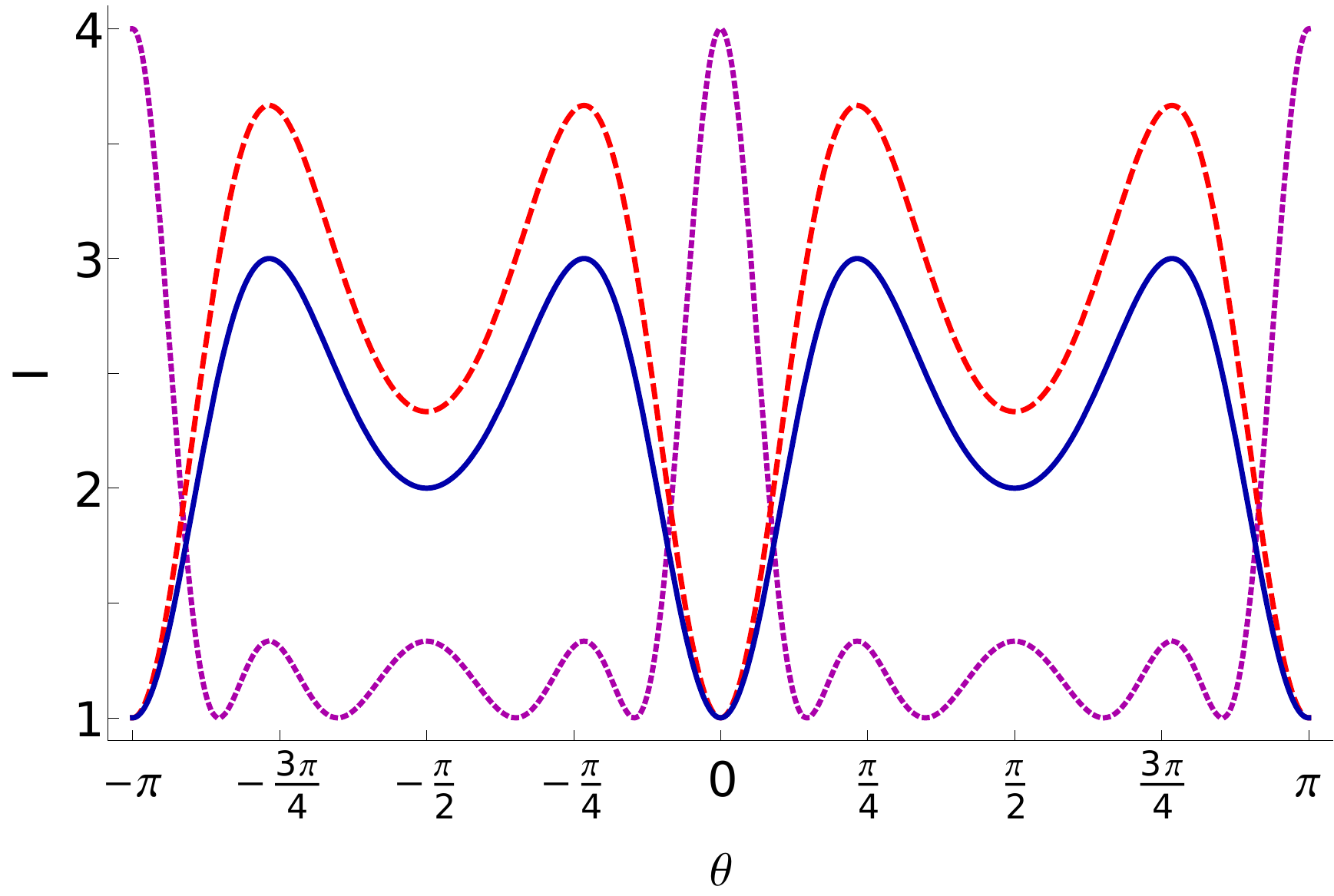}
\caption{\label{PLOTDIR3} (Color online) Plot of the angular dependent intensity of the initial anti-symmetric state $\ket{W^-_{2,1}}$ (striped) and the anti-symmetric state $\ket{\tilde{W}^{-}_{2,1}}$ (solid) compared to the symmetric W-state $\ket{W_{2,1}}$ (dotted); $k\,d = \frac{3}{2}\pi$ and $\mathrm{I}$ is given in units of the intensity produced by a one atom system.}
\end{figure}

\noindent Fig.~\ref{PLOTDIR3} shows that the anti-symmetric states $\ket{W^-_{2,1}}$ (striped) and $\ket{\tilde{W}^{-}_{2,1}}$ (solid) have global and identical minima in the directions $\theta = 0\,,\,\pm\pi$, but differ for $\theta \neq 0\,,\,\pm\pi$. In contrast the corresponding symmetric state (dotted) has global maxima at $\theta = 0\,,\,\pm\pi$, but - what is not intuitive - is subradiant over larger areas than the corresponding anti-symmetric states. This result suggests that one cannot decide whether a system of entangled atoms will show super- or subradiant emission by just considering the initial state - the behavior is rather dependent on the particular direction of observation. Furthermore, the anti-symmetric states $\ket{\tilde{W}^{-}_{2,1}}$ and $\ket{W^-_{2,1}}$ have the same intensity minima (cf.,~Eq.~(\ref{dickesub})). Thus, they are indistinguishable if one were to concentrate on the rate of emission $\theta \neq 0\,,\,\pm\pi$ alone. However, the state $\ket{\tilde{W}^{-}_{2,1}}$ shows a more subradiant behavior for $\theta \neq 0\,,\,\pm\pi$ making a distinction between those two states possible if the angular dependency is taken into account.

We want to note that we have derived the quantum paths and intensity distributions for all anti-symmetric states up to $N=3$ suggesting that our quantum path approach used in this section remains valid also for a broader class of states. 

\section{Conclusion}

In conclusion we studied enhanced spontaneous emission from entangled atoms for generalized symmetric W-states. We traced back the enhancement to  interferences of multiple photon quantum paths and introduced a framework which enables to precisely identify each specific quantum path leading to the enhanced radiation. Our physical framework is valid for states of zero dipole moment, where a classical antenna interpretation for the enhanced radiation is not applicable. Furthermore we investigated the angular distribution of the emission of photons by entangled atoms and showed strong focussing of light emitted by entangled atoms. We extended our investigation to non-symmetric generalized W-states and gave examples which support our interpretation in terms of interference even if we consider a broader class of states. Finally, we showed that also symmetric W-states can emit in a subradiant manner. This underlines the importance of considering position dependent detection to fully describe the radiation properties of entangled atoms. While we have concentrated here on the mean intensity of the emitted radiation it would be worthwhile to investigate other quantum signatures of entangled states in the emitted radiation.

\section{Ackowledgements}

GSA thanks M.~O.~Scully and J.~Dowling for interesting discussions on this paper. JvZ thanks the Erlangen Graduate School in Advanced Optical Technologies (SAOT) by the German Research Foundation (DFG) in the framework of the German excellence initiative for funding. RW~gratefully acknowledges financial support by the Elite Network of Bavaria and the hospitality at the Oklahoma State University in spring 2010 and 2011. This work was supported by the DFG.

\appendix

\section{Derivation of the angular dependent distribution of intensity for initial generalized W-states}
\label{app1}

In this appendix we want to derive Eq.~(\ref{IWPHI}) from first principles. Starting with Eq.~(\ref{IGEN}) the angular dependent intensity from generalized symmetric W-states reads

\staffeld
\label{app1eq1}
\mathrm{I}_{\ket{W_{n_e,n_g}}} &=& \sum_{i,j} \langle \hat{s}^{+}_i \hat{s}^{-}_j \rangle \,e^{i\,(\varphi_i-\varphi_j)} \nonumber\\
&=&\sum_{i} \langle \hat{s}^{+}_i \hat{s}^{-}_i \rangle + \sum_{i \neq j} \langle \hat{s}^{+}_i \hat{s}^{-}_j \rangle \,e^{i\,(\varphi_i-\varphi_j)} \nonumber\\
&=& N\,\alpha + \beta \left( \sum_{i,j} e^{i\,(\varphi_i-\varphi_j)} - N \right) \nonumber \\
&=& N\,\alpha + \beta \left|\left|\sum_i e^{i\varphi_i}\right|\right|^2 - \beta\,N  \nonumber \\
&=& N\,\alpha + \beta \frac{\sin^2(\frac{\varphi_N}{2})}{\sin^2(\frac{\varphi}{2})} - \beta\,N \,,
\stoffeld

\noindent where we used the fact that due to the symmetry of the W-states the matrix elements $\langle \hat{s}^{+}_i \hat{s}^{-}_i \rangle$ and $\langle \hat{s}^{+}_i \hat{s}^{-}_j \rangle$ are independent of $i,j$ and given by constants which we denote by $\alpha$ and $\beta$, respectively. For $\alpha$ we obtain

\staffeld
\langle \hat{s}^{+}_i \hat{s}^{-}_i \rangle \equiv \alpha &=& \frac{1}{2} + \langle \hat{s}^{z}_i \rangle \nonumber \\
\rightarrow N\,\alpha &=& \frac{N}{2} + \langle \hat{s}^{z} \rangle = \frac{N}{2} + m = n_e \,,
\stoffeld

\noindent as $m = n_e - N/2$ (cf.~section~\ref{DickeSEC}). Now we have to determine the constant $\beta$. The sum over all matrix elements $\langle s_i^+ s_j^- \rangle$ of the collective states $\ket{N/2,m}$, corresponding to the symmetric W-states $\ket{W_{n_e,n_g}}$, calculates to \cite{Dicke54}

\staffeld
  \sum_{i,j}\langle s_i^+ s_j^- \rangle &=& \frac{N}{2}(\frac{N}{2} + 1) - m^2 + m \nonumber\\
&=& n_e ( N - n_e -1) 
\stoffeld

\noindent and must be identical to the maximum intensity from generalized symmetric W-states (cf.~Eq.~(\ref{app1eq1})). Thus it follows that

\staffeld
&&n_e ( N - n_e -1) \equiv n_e + \beta\,N\,(N-1) \nonumber \\ 
&&\rightarrow \beta = \frac{n_e ( N - n_e)}{N ( N -1)}\;,
\stoffeld

\noindent as the maximum of $\frac{\sin^2(\frac{\varphi_N}{2})}{\sin^2(\frac{\varphi}{2})} = N^2$. Now, if we put $\alpha$ and $\beta$ into Eq.~\ref{app1eq1}, we arrive at the angular dependent intensity from generalized symmetric W-states introduced in Eq.~(\ref{IWPHI}).

\end{document}